\documentclass{article}

\usepackage{physics}
\usepackage{braket}
\usepackage[square,sort,comma,numbers]{natbib}
\usepackage[english]{babel}
\usepackage{graphicx}
\usepackage{float}
\usepackage{titling}
\usepackage{lipsum}
\usepackage{indentfirst}
\usepackage{amsmath}
\usepackage{amssymb}
\usepackage{caption}
\usepackage{sidecap}
\usepackage{mathrsfs}
\usepackage[a4paper, total={5.75in, 9in}]{geometry}

\sidecaptionvpos{figure}{c}

\title{A Hamiltonian-like formalism that treats one spatial coordinate - rather than time - differently}
\author{Sivapalan Chelvaniththilan \\ email: niththilan@gmail.com}
\date{\today}
\begin{document}
\maketitle
\begin{abstract}
The Hamiltonian and Lagrangian formalisms of Qunatum Field Theory (QFT) are equivalent. But while Lorentz invariance can be clearly seen in the Lagrangian formalism, it is not so explicit in the Hamiltonian one. This is because time is treated a little differently from the spatial coordinates in the Hamiltonian formalism. In this paper, I explore whether it is possible to devise another formalism that is just like the Hamiltonian one (with operators and state vectors) but which treats time on an equal footing with two of the spatial coordinates, while the third one is treated differently, the way time is in the usual Hamiltonian formalism. 
\end{abstract}

\section{Introduction}
In the Hamiltonian formalism \cite{zee2010quantum}, a state vector denotes the state of of all the fields at all points in space at a particular point in time. That is, it represents a slice through spacetime at a fixed value of time. Further, any operator in the Hamiltonian formalism is a function of the various fields at all points in space at a particular time. For example, the Hamiltonian operator for a scalar field $\phi$ is given by \cite{mandl2010quantum}

\begin{eqnarray}
H=\int d^3x \frac{1}{2} \left[ \hat{\Pi}^2 + (\nabla\hat{\phi})^2 + m^2\hat{\phi}^2 \right]
\label{eqn:1}
\end{eqnarray}

Here the conjugate momentum field is defined such that it satisfies the operator equation

\begin{eqnarray}
\hat{\Pi}(x,t)=\frac{\partial\hat{\phi}(x,t)}{\partial t}
\label{eqn:2}
\end{eqnarray}

where I have used $(x,t)$ as a shorthand for $(x,y,z,t)$. Since both the above definitions are not Lorentz invariant, this means that the method of doing the calculations does not treat space and time in the same way. But it can be shown that any physical results obtained from this method respect Lorentz invariance. One way of showing this is by using the equivalence of the Hamiltonian formalism to the Lagrangian one, which is explicitly Lorentz invariant. \\

This naturally raises the question, is time the only dimension that can be treated differently in a particular formalism of QFT? Is it possible to devise a formalism in which a state vector represents a slice through spacetime at a fixed value of, say, the z coordinate? Such a vector would contain information about all the fields at all points along a plane perpendicular to the z axis at all values of time. Then, an operator similar to the Hamiltonian operator would describe its evolution with varying values of z rather than t. The purpose of this paper is to show that this is possible. \\

In order to do this, I will use a method similar to Zee's \cite{zee2010quantum} calculations that derive the Lagrangian formalism from the Hamiltonian one. But I will do this calculation the other way, starting from the Lagranian formalism with its Lorentz-invariant action, then eliminating the time derivative operator in this action and introducing the conjugate momenta to the fields and finally arriving at an expression in the Hamiltonian formalism in terms of state vectors. Then I will try to do something similar with the coordinate z instead of t, eliminating the $\partial/\partial z$ operator and arriving at a set of vectors that represent the state at a fixed value of z. \\

Further, while Zee does this calculation for a single particle, I will do something similar for a field.

\section{Deriving the Hamiltonian from the Lagrangian}
The path integral for a free scalar field is 

\begin{eqnarray}
Z=\int D\phi \exp \left\{ i\int d^4x \frac{1}{2} \left[ \left( \frac{\partial\phi}{\partial t} \right)^2 - \left( \frac{\partial\phi}{\partial x} \right)^2 - \left( \frac{\partial\phi}{\partial y} \right)^2 - \left( \frac{\partial\phi}{\partial z} \right)^2 \right]-V(\phi)  \right\}
\label{eqn:3}
\end{eqnarray}

Defining

\begin{eqnarray}
U=\frac{1}{2} \left[ \left( \frac{\partial\phi}{\partial x} \right)^2 + \left( \frac{\partial\phi}{\partial y} \right)^2 + \left( \frac{\partial\phi}{\partial z} \right)^2 \right] +V(\phi) 
\label{eqn:4}
\end{eqnarray}

this becomes

\begin{eqnarray}
Z=\int D\phi \exp \left\{ i\int d^4x \frac{1}{2} \left( \frac{\partial\phi}{\partial t} \right)^2 - U \right\}
\label{eqn:5}
\end{eqnarray}

Let us discretise the time variable and write the integral over t as a sum. If we consider discrete time steps of duration $\epsilon$ the time-derivative can also be expressed in terms of $\epsilon$

\begin{eqnarray}
Z=\int D\phi \exp \left\{ \sum_t i\int d^3x \frac{1}{2} \epsilon \left( \frac{\phi_{t+\epsilon}-\phi_t}{\epsilon} \right)^2 - U\epsilon \right\} \nonumber \\ \nonumber \\
=\int D\phi \prod_t \exp \left\{ i\int d^3x \frac{1}{2} \frac{(\phi_{t+\epsilon}-\phi_t)^2}{\epsilon} - U\epsilon \right\}
\label{eqn:6}
\end{eqnarray}

where $d^3x$ is shorthand for $dxdydz$. Strictly speaking, there should also be an $x$ in the subscript, that is, $(\phi_{t+\epsilon,x}-\phi_{t,x})$ and there should also be subscripts on U. But I will omit them to avoid clutter. To proceed I will make use of the identity

\begin{eqnarray}
\int d\Pi \exp i\left(-\frac{1}{2}\epsilon\Pi^2+b\Pi\right)=\sqrt{\frac{2\pi}{i\epsilon}}\exp\left(i\frac{b^2}{2\epsilon}\right)
\label{eqn:7}
\end{eqnarray}

(This can be easily derived from the Gaussian integral.) At this point $\Pi$ is just a real variable but it will turn out to be the conjugate momentum. It is not to be confused with the similar symbol for product which appears in a larger font. I have used uppercase $\Pi$ to avoid confusion with the constant $\pi$. Modifying the above identity slightly,

\begin{eqnarray}
\int d\Pi \exp i\left(-\sum_x \epsilon^3\left(\frac{1}{2}\epsilon\Pi_x^2-b_x\Pi_x\right)\right)=\left(\frac{2\pi}{i\epsilon^4}\right)^{N/2}\exp\left(i\sum_x \epsilon^3\frac{b_x^2}{2\epsilon}\right)
\label{eqn:8}
\end{eqnarray}

where N is the number of possible values of x. Now in the continuum limit, $\sum_x \epsilon^3$ becomes $\int d^3x$. If we further identify b with $(\phi_{t+\epsilon}-\phi_t)$ then Equation \ref{eqn:6} becomes

\begin{eqnarray}
Z=\int D\phi \prod_t  \left(\prod_x \sqrt{\frac{i\epsilon^4}{2\pi}} \int d\Pi_x \right)  \exp \left\{ -i\int d^3x \left[ \frac{1}{2}\epsilon \Pi^2 - (\phi_{t+\epsilon}-\phi_t)\Pi + U\epsilon \right] \right\}
\label{eqn:9}
\end{eqnarray}

where again I am omitting some of the subscripts for clarity. Now the path integral measure is defined as

\begin{eqnarray}
D\phi =\prod_{x,t} \frac{1}{\sqrt{2\pi i \epsilon^4}}d\phi_{x,t}
\label{eqn:10}
\end{eqnarray}

Substituting this, the path integral becomes

\begin{eqnarray}
Z=\prod_t  \left(\prod_x \frac{1}{2\pi} \int d\phi_x d\Pi_x \right)  \exp \left\{ -i\int d^3x \left[ \frac{1}{2}\epsilon \Pi^2 - (\phi_{t+\epsilon}-\phi_t)\Pi + U\epsilon \right] \right\}
\label{eqn:11}
\end{eqnarray}

Up to this point there are no operators or state vectors in the calculations. $\phi$ and$\Pi$ are just variables. Let us now define a hermitian operator $\hat{\phi}$ such that it has a uniform non-degenerate spectrum of eigenvalues along the whole of the real  number line. Further, let us denote by $|\phi\rangle$ the eigenvector of $\hat{\phi}$ with eigenvalue $\phi$. This means,

\begin{eqnarray}
\hat{\phi}=\int d\phi \phi |\phi\rangle\langle\phi|
\label{eqn:12}
\end{eqnarray}

Let us normalise these eigenvectors such that

\begin{eqnarray}
\langle\phi_1|\phi_2\rangle=\delta(\phi_1-\phi_2)
\label{eqn:13}
\end{eqnarray}

This implies the completeness relation

\begin{eqnarray}
\int d\phi|\phi\rangle\langle\phi|=1
\label{eqn:14}
\end{eqnarray}

Let us also define a set of vectors $|\Pi\rangle$ labeled by the variable $\Pi$ such that

\begin{eqnarray}
\langle\phi|\Pi\rangle=\exp(i\phi\Pi)
\label{eqn:15}
\end{eqnarray}

The above equation gives the components of the vectors $|\Pi\rangle$ in a basis that is defined with the vectors vectors $|\phi\rangle$ as the basis vectors. What is the normalisation condition satisfied by these vectors? We can find this out using

\begin{eqnarray}
\langle\Pi_1|\Pi_2\rangle=\int d\phi \langle\Pi_1|\phi\rangle\langle\phi|\Pi_2\rangle \nonumber \\ \nonumber \\
=\int d\phi \exp[\phi(\Pi_1-\Pi_2)] =2\pi\delta (\Pi_1-\Pi_2)
\label{eqn:16}
\end{eqnarray}

This implies the completeness relation

\begin{eqnarray}
\frac{1}{2\pi}\int d\Pi|\Pi\rangle\langle\Pi|=1
\label{eqn:17}
\end{eqnarray}

Now let us define operators $\hat{\phi}$ and $\hat{\Pi}$ for each point in space. Using the properties of these operators that we just found, Equation \ref{eqn:9} can be written as

\begin{eqnarray}
Z=\prod_t  \left(\prod_x \frac{1}{2\pi} \int d\phi_x d\Pi_x \right)  \exp \left\{ -i\int d^3x \frac{1}{2}\epsilon \Pi^2 + U\epsilon \right\} \langle\phi_{t+\epsilon}|\Pi_x\rangle\langle\Pi_x|\phi_t\rangle
\label{eqn:18}
\end{eqnarray}

Note that since we have defined one operator $\hat{\phi}$ for each point in space, rather than each point in spacetime, this means we are in the Schrodinger picture and not in the Heisenberg picture (in which the operators depend on time). Since the operator $\hat{\phi}$ does not change with time, neither do its eigenstates $|\phi\rangle$ and $\langle\phi|$ The subscripts such a $t$ and $t+\epsilon$ on the eigenstates simply mean that they are the eigenstates that the system is in at the times $t$ and $t+\epsilon$. They do not mean that they are eigenstates of two different operators defined at $t$ and $t+\epsilon$. (That would be the Heinsenberg picture.) Later in this paper, I will also consider the properties of the operators in the Heinsenberg picture. \\

Now the terms involving $\Pi$ and $U$ (which is a function of $\phi$) can be replaced by the corresponding operators if they are placed in front of their relevant eigenvectors.

\begin{eqnarray}
Z=\prod_t  \left(\prod_x \frac{1}{2\pi} \int d\phi_x d\Pi_x \right) \times \nonumber \\ \nonumber \\
\langle\phi_{t+\epsilon}|  \exp \left\{ -i\int d^3x \frac{1}{2}\epsilon \hat{\Pi}^2 \right\}    |\Pi_x\rangle\langle\Pi_x|  \exp\left\{  -i\int d^3x\hat{U}\epsilon \right\} | \phi_t\rangle
\label{eqn:19}
\end{eqnarray}

Using the completeness relation, this can be simplified to give

\begin{eqnarray}
Z=\prod_t  \left(\prod_x \int d\phi_x \right)  \langle\phi_{t+\epsilon}|  \exp \left\{ -i\int d^3x \frac{1}{2}\epsilon \hat{\Pi}^2 \right\} \exp\left\{  -i\int d^3x\hat{U}\epsilon \right\} | \phi_t\rangle
\label{eqn:20}
\end{eqnarray}

Usually we cannot combine the two exponentials because the commutator of $\Pi$ and $U$ would give extra terms, but in this case since $\epsilon$ is small and because the terms involving the commutator will be second order in $\epsilon$, they will vanish when we take the continuum limit in which $\epsilon$ tends to zero. Hence

\begin{eqnarray}
Z=\prod_t  \left(\prod_x \int d\phi_x \right)  \langle\phi_{t+\epsilon}|  \exp \left\{ -i\hat{H}\epsilon \right\} | \phi_t\rangle
\label{eqn:21}
\end{eqnarray}

where

\begin{eqnarray}
\hat{H}=\int d^3x \left\{\frac{1}{2} \hat{\Pi}^2 +\hat{U} \right\}
\label{eqn:22}
\end{eqnarray}

Now if we expand the product in Equation \ref{eqn:21} and use the completeness relation of $\phi$ to simplify this becomes

\begin{eqnarray}
Z=\langle\phi_{final}|  \exp \left\{ -i\hat{H} ~\Delta t \right\} | \phi_{initial}\rangle
\label{eqn:23}
\end{eqnarray}

where $| \phi_{initial}\rangle$ and $| \phi_{final}\rangle$ are the states of the fields at two hypersurfaces at two different values of t which form the boundary of the region over which the integral of the Lagrangian is taken. $ ~\Delta t$ is the time interval between these two states. Thus the Hamiltonian formalism can be derived from the Lagrangian one.

\section{Deriving a Hamiltonian-like operator for the z coordinate}
The above calculations treated the time coordinate t differently to the other three. Now let us perform a similar calculation that treats z differently. In order to do this I will define the quantity $U'$ in a similar way to $U$ defined in Equation \ref{eqn:4}

\begin{eqnarray}
U'=\frac{1}{2} \left[ \left( \frac{\partial\phi}{\partial x} \right)^2 + \left( \frac{\partial\phi}{\partial y} \right)^2 - \left( \frac{\partial\phi}{\partial t} \right)^2 \right] +V(\phi) 
\label{eqn:24}
\end{eqnarray}

Using this the path integral can be written as

\begin{eqnarray}
Z=\int D\phi \exp \left\{ i\int d^4x \left[- \frac{1}{2} \left( \frac{\partial\phi}{\partial z} \right)^2 - U' \right] \right\}
\label{eqn:25}
\end{eqnarray}

Discretising the z coordinate gives us

\begin{eqnarray}
Z=\int D\phi \prod_z \exp \left\{ i\int d^3x' \left[-\frac{1}{2} \frac{(\phi_{z+\epsilon}-\phi_z)^2}{\epsilon} - U'\epsilon \right]  \right\}
\label{eqn:26}
\end{eqnarray}

Here, I am using $d^3x'$ as a shorthand for $dxdydt$ (while $d^3x$, without the prime, is a shorthand for $dxdydz$). Now let us use an identity similar to Equation \ref{eqn:8} with some sign differences.

\begin{eqnarray}
\int d\Pi \exp i\left(-\sum_{x'} \epsilon^3\left(-\frac{1}{2}\epsilon\Pi_{x'}^2+b_x'\Pi_{x'}\right)\right)=\left(\frac{2\pi}{-i\epsilon^4}\right)^{N/2}\exp\left(-i\sum_{x'} \epsilon^3\frac{b_{x'}'^{2}}{2\epsilon}\right)
\label{eqn:27}
\end{eqnarray}

where again the subscript $x'$ on a sum or product implies that it must be performed over all values of x, y and t in the discretised spacetime, i.e. over a hyperplane with a fixed value of z. Note that there is a minus sign in front of the $i\epsilon$ unlike in Equation \ref{eqn:8} but we can ignore this by taking N/2 to be an even number (which we can since we later take the limit $N\rightarrow\infty$) but even if we don't, this will only give us an overall phase. When we substitute $(\phi_{z+\epsilon}-\phi_z)$ for $b'$, this identity enables us to write Equation \ref{eqn:26} as

\begin{eqnarray}
Z=\int D\phi \prod_z  \left(\prod_{x'} \sqrt{\frac{i\epsilon^4}{2\pi}} \int d\Pi_{x'} \right)  \exp \left\{ -i\int d^3x' \left[ -\frac{1}{2}\epsilon \Pi^2 - (\phi_{z+\epsilon}-\phi_z)\Pi + U'\epsilon \right] \right\}
\label{eqn:28}
\end{eqnarray}

Now let us again define operators $\hat{\phi}$ and $\hat{\Pi}$ in a similar way to the previous section, but instead of defining them at every point in space (which is the set of all points with coordinates of the form (x,y,z) ) let us define them at every point in the set with coordinates (x,y,t). By analogy with Equation \ref{eqn:18} this gives us

\begin{eqnarray}
Z=\prod_z  \left(\prod_{x'} \frac{1}{2\pi} \int d\phi_{x'} d\Pi_{x'} \right)  \exp \left\{ -i\int d^3x' \left[ -\frac{1}{2}\epsilon \Pi^2 + U'\epsilon \right] \right\} \langle\phi_{z+\epsilon}|\Pi_{x'}\rangle\langle\Pi_{x'}|\phi_z\rangle
\label{eqn:29}
\end{eqnarray}

Following through the calculations of Equations \ref{eqn:18} to \ref{eqn:23} leads to

\begin{eqnarray}
Z=\langle\phi_{final}|  \exp \left\{ -i\hat{H'}  ~\Delta z \right\} | \phi_{initial}\rangle
\label{eqn:30}
\end{eqnarray}

where the initial and final states are defined on hyperplanes of constant values of the z coordinate and $ ~\Delta z$ is the distance between those hyperplanes. The operator $\hat{H'}$ is defined as

\begin{eqnarray}
\hat{H'}=\int d^3x' \left\{-\frac{1}{2}\hat{\Pi}^2 +\hat{U'} \right\}
\label{eqn:31}
\end{eqnarray}

\section{Properties of the conjugate momentum}
Since the Hamiltonian and the analogous operator in the new formalism are defined in terms of both the field and its conjugate momentum, I will now derive some useful properties of it. For each property, I will first derive it in the usual Hamiltonian formalism and then in the new formalism. For these, I will use Equations \ref{eqn:12} to \ref{eqn:17} which are true in both formalisms. Let us start with the commutation relation of the field with its momentum. Most textbooks \cite{zee2010quantum} \cite{mandl2010quantum} present these commutation relations as simply assumptions. But it is possible to prove them mathematically using the definitions of these operators.

\subsection{Commutation relations}

First, since there is a $\hat{\phi}_x$ and a $\hat{\Pi}_x$ operator for every point $x$ in space, let us denote by $|\phi\rangle$ the state of the field which is an eigenstate of all the $\hat{\phi}$ operators. Let us define $|\Pi\rangle$ similarly. Then the two operators at each point are

\begin{eqnarray}
\hat{\phi}_x=\int d^N\phi ~\phi_x |\phi\rangle\langle\phi|
\label{eqn:32}
\end{eqnarray}

\begin{eqnarray}
\hat{\Pi}_x=\frac{1}{(2\pi)^N} \int d^N\Pi ~\Pi_x ~|\Pi\rangle\langle\Pi|
\label{eqn:33}
\end{eqnarray}

where N is the number of points in space considered in the discretised model. Now a general state can be expressed in the eigenbasis of the $\hat{\phi}$ operators as

\begin{eqnarray}
|f\rangle=\int d^N\phi ~f(\phi) |\phi\rangle
\label{eqn:34}
\end{eqnarray}

where $f(\phi)$ is shorthand for $f(\phi_1,\phi_2,\cdots,\phi_N)$. When one of the momentum operators acts on this state,

\begin{eqnarray}
\hat{\Pi}_x|f\rangle=\frac{1}{(2\pi)^N} \int d^N\phi ~d^N\Pi ~\Pi_x f(\phi) |\Pi\rangle\langle\Pi|\phi\rangle
\label{eqn:35}
\end{eqnarray}

To simplify, we will insert into this the expansion of the identity

\begin{eqnarray}
1=\int d^N\phi'~ |\phi'\rangle\langle\phi'|
\label{eqn:36}
\end{eqnarray}

to get

\begin{eqnarray}
\hat{\Pi}_x|f\rangle=\frac{1}{(2\pi)^N} \int d^N\phi ~d^N\phi' ~d^N\Pi ~\Pi_x f(\phi) |\phi'\rangle\langle\phi'|\Pi\rangle\langle\Pi|\phi\rangle \nonumber \\ \nonumber \\
=\frac{1}{(2\pi)^N} \int d^N\phi ~d^N\phi'~ d^N\Pi ~ f(\phi) |\phi'\rangle \bigg[ \Pi_x \exp\left( i\phi'\Pi - i\phi\Pi \right) \bigg] \nonumber \\ \nonumber \\
=\frac{1}{(2\pi)^N} \int d^N\phi~ d^N\phi'~ f(\phi) |\phi'\rangle \left(i\frac{\partial}{\partial \phi_x}\right) \int d^N\Pi \exp\left( i\phi'\Pi - i\phi\Pi \right) \nonumber \\ \nonumber \\
=\frac{1}{(2\pi)^N} \int d^N\phi~ d^N\phi'~ f(\phi) |\phi'\rangle \left(i\frac{\partial}{\partial \phi_x} (2\pi)^N \delta^N\left( \phi' - \phi \right) \right) \nonumber \\ \nonumber \\
=\int d^N\phi ~d^N\phi' |\phi'\rangle \left(-i\frac{\partial f(\phi)}{\partial \phi_x}\right) \delta^N\left( \phi' - \phi \right) \nonumber \\ \nonumber \\
\therefore\hat{\Pi}_x|f\rangle=\int d^N\phi ~ \left(-i\frac{\partial f(\phi)}{\partial \phi_x}\right) |\phi\rangle
\label{eqn:37}
\end{eqnarray}

~\\
In the above, $\phi\Pi$ is shorthand for $\phi_1\Pi_1+\phi_2\Pi_2+\cdots+\phi_N\Pi_N$ and can be thought of as a dot product of the vectors $(\phi_1,\phi_2,\cdots,\phi_N)$ and $(\Pi_1,\Pi_2,\cdots,\Pi_N)$ which consist of the values of the field and its conjugate momentum at every point in space. The result obtained above implies that the conjugate momentum can be expressed as

\begin{eqnarray}
\hat{\Pi}_x=-i\frac{\partial}{\partial \phi_x}
\label{eqn:38}
\end{eqnarray}

which reminds us of the similar expression for the position and momentum of a particle in quantum mechanics, which can be proved in a similar way. In the continuum limit, the partial derivative is replaced by a functional derivative

\begin{eqnarray}
\hat{\Pi}(x)=-i\frac{\delta}{\delta \phi(x)}
\label{eqn:39}
\end{eqnarray}

Further, in the continuum limit, the state of the field is represented by a wavefunctional (rather than a wavefunction) $\Psi$. Acting on this wavefunctional with $\hat{\Pi}(x_1)\hat{\phi}(x_2)$,

\begin{eqnarray}
\hat{\Pi}(x_1)\hat{\phi}(x_2)\Psi[\phi]=-i\frac{\delta}{\delta \phi(x_1)}\bigg[ \phi(x_2)\Psi[\phi] \bigg] \nonumber \\ \nonumber \\
=-i\delta(x_1-x_2) \Psi[\phi] -i \phi(x_2) \frac{\delta}{\delta \phi(x_1)}\bigg[\Psi[\phi] \bigg] \nonumber \\ \nonumber \\
=-i\delta(x_1-x_2) \Psi[\phi] + \hat{\phi(x_2)} \hat{\Pi(x_1)} \Psi[\phi]  \nonumber \\ \nonumber \\
\therefore \hat{\Pi}(x_1)\hat{\phi}(x_2)-\hat{\phi}(x_2) \hat{\Pi}(x_1)=-i\delta(x_1-x_2)
\label{eqn:40}
\end{eqnarray}

Here $\Psi[\phi]$ is shorthand for $\Psi[\phi(x_1),\phi(x_2),\cdots ,\phi(x_N)]$, that is, $\Psi$ is a function of the values of $\phi$ at each point in space. In the continuum limit, this means $\Psi$ is a functional of $\phi$. \\

This commutation relation is one of the main results of quantum field theory. Equations \ref{eqn:39} and \ref{eqn:40} also hold in the new formalism with $x$ replaced by $x'$. (Here $x$ is used as a shorthand for the coordinates $(x,y,z)$ and $x'$ for $(x,y,t)$ )

\subsection{Time-evolution or z-evolution relations}
In the usual Hamiltonian formalism, to find the time evolution equation, we find the commutator of the field with the Hamiltonian

\begin{eqnarray}
[\hat{H},\hat{\phi}(x_1)]= \left[ \int d^3x_2 \left\{\frac{1}{2} \left(\hat{\Pi}(x_2)\right)^2 +\hat{U}(x_2) \right\}, \hat{\phi}(x_1) \right] \nonumber \\ \nonumber \\
= \frac{1}{2} \int d^3x_2  \left[  \left(\hat{\Pi}(x_2)\right)^2, \hat{\phi}(x_1) \right] \nonumber \\ \nonumber \\
=\frac{1}{2} \int d^3x_2 ~ 2\hat{\Pi}(x_2)(-i)\delta(x_2-x_1)=-i\hat{\Pi}(x_1)
\label{eqn:41}
\end{eqnarray}

This implies that in the Heinsenberg picture,

\begin{eqnarray}
\frac{\partial\hat{\phi}}{\partial t}=i[\hat{H},\hat{\phi}]=\hat{\Pi}
\label{eqn:42}
\end{eqnarray}

Similarly in the new formalism,

\begin{eqnarray}
[\hat{H'},\hat{\phi}(x_1)]= \left[ \int d^3x_2 \left\{-\frac{1}{2} \left(\hat{\Pi}(x_2)\right)^2 +\hat{U'}(x_2) \right\}, \hat{\phi}(x_1) \right] \nonumber \\ \nonumber \\
= -\frac{1}{2} \int d^3x_2  \left[  \left(\hat{\Pi}(x_2)\right)^2, \hat{\phi}(x_1) \right] \nonumber \\ \nonumber \\
-\frac{1}{2} \int d^3x_2 ~ 2\hat{\Pi}(x_2)(-i)\delta(x_2-x_1)=i\hat{\Pi}(x_2)
\label{eqn:43}
\end{eqnarray}

Hence

\begin{eqnarray}
\frac{\partial\hat{\phi}}{\partial z}=i[\hat{H'},\hat{\phi}]=-\hat{\Pi}
\label{eqn:44}
\end{eqnarray}

We can use a similar method to find $\partial\hat{\Pi}/\partial z$ an combine both the relations to get the equation of motion. But this won't be necessary since we know that both the formalisms are equivalent and must result in the same equation of motion, which in the case of a free field (where $V(\phi)=m^2\hat{\phi}^2/2$) is

\begin{eqnarray}
\frac{\partial^2\hat{\phi}}{\partial t^2}-\frac{\partial^2\hat{\phi}}{\partial x^2}-\frac{\partial^2\hat{\phi}}{\partial y^2}-\frac{\partial^2\hat{\phi}}{\partial z^2}+m^2\hat{\phi}^2=0
\label{eqn:45}
\end{eqnarray}

Note that in both formalisms, the evolution of state vectors in the Schrodinger picture has the same form (Equations \ref{eqn:23} and \ref{eqn:30}). So in the former, when we change to the Heinsenberg picture we get, for any operator $\hat{A}$,

\begin{eqnarray}
\hat{A}(t)=\exp(i\hat{H}t)\hat{A}(0)\exp(-i\hat{H}t) \nonumber \\ \nonumber \\
\therefore\frac{\partial\hat{A}}{\partial t}=i[\hat{H},\hat{A}]
\label{eqn:46}
\end{eqnarray} 

whereas in the latter we get

\begin{eqnarray}
\hat{A}(z)=\exp(i\hat{H}'z)\hat{A}(0)\exp(-i\hat{H}'z) \nonumber \\ \nonumber \\
\therefore\frac{\partial\hat{A}}{\partial z}=i[\hat{H}',\hat{A}]
\label{eqn:47}
\end{eqnarray} 

\section{The creation and annihilation operators}
The expansion of the field in terms of the creation and annihilation operators and the derivation of their commutaton relations are provided in several standard texts such as \cite{zee2010quantum} and \cite{mandl2010quantum}. The field can be expressed in the Heinsenberg picture as (from this point I omit the hats on operators to avoid clutter)

\begin{eqnarray}
\phi(x,y,z,t)=\int\frac{dk_xdk_ydk_z}{\sqrt{(2\pi)^3 2\omega}}\bigg[ a(\overrightarrow{k})\exp{-i(\omega t-k_xx-k_yy-k_zz)} \nonumber \\ \nonumber \\
+a^\dagger (\overrightarrow{k})\exp{i(\omega t-k_xx-k_yy-k_zz)} \bigg]
\label{eqn:48}
\end{eqnarray} 

This is essentially a Fourier expansion of the field $\phi$. The equation of motion implies that

\begin{eqnarray}
\omega(\overrightarrow{k})=\sqrt{k_x^2+k_y^2+k_z^2+m^2}
\label{eqn:49}
\end{eqnarray} 

and the hermiticity of the field implies that the operators denoted by $a$ and $a^\dagger$ must be hermitian conjugates of each other and hence the notation. To get the expression for $\phi$ in the Schrodinger picture we can simply ignore the variation in $t$, i.e. omit the $\pm\omega t$ terms in the exponentials. Now let us try a similar expansion in the new formalism. 

\subsection{Hermitian conjugation relations}
Now the integration must be over the variables $k_x$, $k_y$ and $k_t$ rather than $k_x$, $k_y$ and $k_z$. Hence we get

\begin{eqnarray}
\phi(x,y,z,t)=\int\frac{dk_xdk_ydk_t}{\sqrt{(2\pi)^3 2|\lambda|}}\bigg[ a'(\overrightarrow{k'})\exp{i(\lambda z+k_xx+k_yy-k_tt)} \nonumber \\ \nonumber \\
+\overline{a'} (\overrightarrow{k'})\exp{-i(\lambda z+k_xx+k_yy-k_tt)} \bigg]
\label{eqn:50}
\end{eqnarray} 

Here I have chosen the signs inside the exponential to be the same as in Equation \ref{eqn:48}. But it is also possible to choose them differently, for example to choose them such that the exponential is $\exp{-i(\lambda z-k_xx-k_yy-k_tt)}$ This would simply mean defining $k_x$, $k_y$, $k_z$ and $\lambda$ differently. \\

Here I use the symbol $\overrightarrow{k'}$ to mean the vector $(k_x,k_y,k_t)$ while $k$ without the prime means $(k_x,k_y,k_z)$. Unlike $\omega$ in the usual Hamiltonian formalism, $\lambda$ which is defined as

\begin{eqnarray}
\lambda(\overrightarrow{k'})=\sqrt{k_t^2-k_x^2-k_y^2-m^2}
\label{eqn:51}
\end{eqnarray}

could be real or imaginary. This means the operator $\phi$ cannot be hermitian for all values of z. But we defined it as having real eigenvalues. I will explain in Section 7 below how these two statements can be consistent with one another. For the moment, I will just note that we can take $\phi$ to be hermitian for $z=0$ because then $(\lambda z+k_xx+k_yy-k_tt)$ will be a real number irrespective of the value of $\lambda$. \\

Further, because $\lambda$ could be imaginary, the hermiticity of $\phi$ and $\Pi$ at $z=0$ does not necessarily guarantee that that the operators $a'$ and $\overline{a'}$ are hermitian conjugates of one another and hence the change in notation, using a bar instead of the dagger symbol. Let us examine the relation between these operators more closely. For this let us divide the $(k_x,k_y,k_t)$ hyperplane into two regions, $P_1$ and $P_2$ where $\lambda$ is real or imaginary respectively. The points in $P_1$ are the ones that satisfy

\begin{eqnarray}
k_t^2\ge k_x^2+k_y^2+m^2
\label{eqn:52}
\end{eqnarray}

and the points in $P_2$ are the ones that don't. The boundary between these two regions is a hyperboloid. For any vector $\overrightarrow{k'}=(k_x,k_y,k_t)$ that lies in the region $P_1$, the operators $a'$ and $\overline{a'}$ are indeed hermitian conjugates of  each other. This can be proved as follows. $\phi$, and similarly $\Pi$ are hermitian at $z=0$. Let us now define the functions $\widetilde{\phi}$ and $\widetilde{\Pi}$ as the ones obtained by Fourier transforming $\phi$ and $\Pi$ in just the x, y and t coordinates (but not in z) and then setting $z=0$. These functions must satisfy

\begin{eqnarray}
\widetilde{\phi}(k_x,k_y,k_t)^\dagger=\widetilde{\phi}(-k_x,-k_y,-k_t) \nonumber \\ \nonumber \\
\widetilde{\Pi}(k_x,k_y,k_t)^\dagger=\widetilde{\Pi}(-k_x,-k_y,-k_t)
\label{eqn:53}
\end{eqnarray}

This is a standard result about the Fourier transforms of Hermitian operators. But performing the Fourier transform on Equation \ref{eqn:50} gives

\begin{eqnarray}
\widetilde{\phi}(k_x,k_y,k_t)=\frac{1}{\sqrt{2|\lambda|}}\left[a'(k_x,k_y,-k_t)+\overline{a'}(-k_x,-k_y,k_t) \right]
\label{eqn:54}
\end{eqnarray}

and differentiating Equation \ref{eqn:50} and using Equation \ref{eqn:44} we get the expression for the momentum operator

\begin{eqnarray}
\Pi(x,y,z,t)=-\int\frac{i\lambda~dk_xdk_ydk_t}{\sqrt{(2\pi)^3 2|\lambda|}}\bigg[ a'(\overrightarrow{k'})\exp{i(\lambda z+k_xx+k_yy-k_tt} \nonumber \\ \nonumber \\
-\overline{a'} (\overrightarrow{k'})\exp{-i(\lambda z+k_xx+k_yy-k_tt)} \bigg]
\label{eqn:55}
\end{eqnarray} 

from which we get

\begin{eqnarray}
\widetilde{\Pi}(k_x,k_y,k_t)=\frac{-i\lambda}{\sqrt{2|\lambda|}}\left[a'(k_x,k_y,-k_t)-\overline{a'}(-k_x,-k_y,k_t)\right]
\label{eqn:56}
\end{eqnarray}

Combining Equations \ref{eqn:54} and \ref{eqn:56} 

\begin{eqnarray}
\sqrt{\frac{|\lambda|}{2}}\left[ \widetilde{\phi}(k_x,k_y,k_t)-\frac{\widetilde{\Pi}(k_x,k_y,k_t)}{i\lambda} \right]=a'(k_x,k_y,-k_t)
\label{eqn:57}
\end{eqnarray}

\begin{eqnarray}
\sqrt{\frac{|\lambda|}{2}}\left[ \widetilde{\phi}(k_x,k_y,k_t)+\frac{\widetilde{\Pi}(k_x,k_y,k_t)}{i\lambda} \right]=\overline{a'}(-k_x,-k_y,k_t)
\label{eqn:58}
\end{eqnarray}

Using Equations \ref{eqn:58} and \ref{eqn:57} and their complex conjugates we can see that if $\lambda$ is real then

\begin{eqnarray}
\left[a'(k_x,k_y,-k_t)\right]^\dagger=\overline{a'}(k_x,k_y,-k_t) \nonumber \\ \nonumber \\
\therefore [a'(k_x,k_y,k_t)]^\dagger=\overline{a'}(k_x,k_y,k_t)
\label{eqn:59}
\end{eqnarray}

which has the same form as the corresponding relations in the usual Hamiltonian formalism. But if $\lambda$ is imaginary, that is if the momentum-space point we are considering is in the region $P_2$, we can use Equation \ref{eqn:53} to show that

\begin{eqnarray}
~~~~~~~~[a'(k_x,k_y,k_t)]^\dagger=a'(-k_x,-k_y,-k_t) \nonumber \\ \nonumber \\
\mathrm{and}~~~[\overline{a'}(k_x,k_y,k_t)]^\dagger=\overline{a'}(-k_x,-k_y,-k_t)
\label{eqn:60}
\end{eqnarray}

which are very different relations. 

\subsection{Commutation relations}
Next, let us find the commutation relations between the creation and annihilation operators. In section 4.1 we showed that

\begin{eqnarray}
[ \Pi(x_1,y_1,t_1),\phi(x_2,y_2,t_2)]=-i\delta(x_1-x_2)\delta(y_1-y_2)\delta(t_1-t_2)
\label{eqn:61}
\end{eqnarray}

This equation was derived in the Schrodinger picture but by multiplying both sides by $e^{-iH'z}$ on the right and $e^{-iH'z}$ on the left we can see that it is also true in the Heisenberg picture. In a similar way it can also be shown that

\begin{eqnarray}
\left[\Pi(x_1,y_1,t_1),\Pi(x_2,y_2,t_2)\right]=0 \nonumber \\ \nonumber \\
\left[\phi(x_1,y_1,t_1),\phi(x_2,y_2,t_2)\right]=0 
\label{eqn:62}
\end{eqnarray}

Expressing these fields in terms of their Fourier transforms and simplifying leads to

\begin{eqnarray}
\left[\widetilde{\Pi}(k_{x1},k_{y1},k_{t1}),\widetilde{\phi}(k_{x2},k_{y2},k_{t2})\right]=-i\delta(k_{x1}+k_{x2})\delta(k_{y1}+k_{y2})\delta(k_{t1}+k_{t2}) \nonumber \\ \nonumber \\
\left[\widetilde{\Pi}(k_{x1},k_{y1},k_{t1}),\widetilde{\Pi}(k_{x2},k_{y2},k_{t2})\right]=0 ~~~~~~~~~~~~~~~~~~~~~~ \nonumber \\ \nonumber \\
\left[\widetilde{\phi}(k_{x1},k_{y1},k_{t1}),\widetilde{\phi}(k_{x2},k_{y2},k_{t2})\right]=0 ~~~~~~~~~~~~~~~~~~~~~~
\label{eqn:63}
\end{eqnarray}

Substituting these into Equations \ref{eqn:57} and \ref{eqn:58} (and renaming some variables in the arguments) gives

\begin{eqnarray}
\left[a'(k_{x1},k_{y1},k_{t1}),\overline{a'}(k_{x2},k_{y2},k_{t2})\right]=\bigg(\frac{|\lambda|}{\lambda}\bigg)\delta(k_{x1}-k_{x2})\delta(k_{y1}-k_{y2})\delta(k_{t1}-k_{t2})  \nonumber \\ \nonumber \\
\left[a'(k_{x1},k_{y1},k_{t1}),a'(k_{x2},k_{y2},k_{t2})\right]=0 ~~~~~~~~~~~~~~~~~~~~~~  \nonumber \\ \nonumber \\
\left[\overline{a'}(k_{x1},k_{y1},k_{t1}),\overline{a'}(k_{x2},k_{y2},k_{t2})\right]=0 ~~~~~~~~~~~~~~~~~~~~~~ 
\label{eqn:64}
\end{eqnarray}

So we can see that when ${\lambda}$ is real the operators $a'$ and $\overline{a'}$ have the same kind of commutation relations as in the usual Hamiltonian formalism but with the opposite sign. \\

Next let us express the Hamiltonian-like operator in terms of these creation and annihilation operators by substituting Equations \ref{eqn:50} and \ref{eqn:55} in Equation \ref{eqn:31}

\begin{eqnarray}
H'=-\int dk_xdk_ydk_t \bigg(\frac{\lambda^2}{2|\lambda|}\bigg) \bigg( a'(k_x, k_y, k_t)\overline{a'}(k_x, k_y, k_t)+\overline{a'}(k_x, k_y, k_t)a'(k_x, k_y, k_t) \bigg)  \nonumber \\ \nonumber \\
=-\int d^3k' \bigg(\frac{\lambda^2}{|\lambda|}\bigg) \bigg(\overline{a'}(\overrightarrow{k'})a'(\overrightarrow{k'}) \bigg)+E'_0~~~~~~~~~~~~~~~~~~~~~~ 
\label{eqn:65}
\end{eqnarray}

where $E'_0$ is the analog of the zero point energy. This implies that

\begin{eqnarray}
\left[H',a'(k_x, k_y, k_t)\right]=\lambda a'(k_x, k_y, k_t) \nonumber \\ \nonumber \\
\left[H',\overline{a'}(k_x, k_y, k_t)\right]=-\lambda \overline{a'}(k_x, k_y, k_t)
\label{eqn:66}
\end{eqnarray}

These can be used to find the z-evolution of the creation and annihilation operators in the Heisenberg picture.

\section{The Feynmann Propagator}
In the usual Hamiltonian formalism, we arrive at the Feynmann propagator by considering a time-ordered product of two fields. So let us see if we can do the same in the new formalism by using a z-ordered product instead. But here is something that we need to be careful about. The time ordered product is sandwiched between an initial and a final vacuum states and the vacuum state is defined as one which gives zero when multiplied by any annihilation operator, i.e. $a(\overrightarrow{k})|0\rangle=0$ which implies $\langle0|a^\dagger(\overrightarrow{k})=0$. Both of these relations are important in simplifying expression involving vacuum states. \\

Suppose we define a similar state $|0'\rangle$ in our new formalism by $a'(\overrightarrow{k'})|0'\rangle=0$. Then when $\overrightarrow{k'}$ is in the region $P_1$ of the $(x,y,t)$ hyperplane, this implies $\langle0'|\overline{a}(\overrightarrow{k})=0$ while if it is in the region $P_2$ it implies that $\langle0'|a(\overrightarrow{k})=0$ ~~(These can be seen by taking the hermitian conjugate using the relations derived in Section 5.1 and I am omitting the minus sign in front of $\overrightarrow{k'}$ here because this equation has to be true for all $\overrightarrow{k'}$) \\

Let us define another state $|0''\rangle$ by $\langle0''|\overline{a}(\overrightarrow{k})=0$ for all $\overrightarrow{k'}$. Then when $\overrightarrow{k'}$ is in $P_1$, this implies that $a'(\overrightarrow{k'})|0''\rangle=0$ and when it is in the region $P_2$ then $\overline{a'}(\overrightarrow{k'})|0''\rangle=0$. Note that both these states behave in the same way in the region $P_1$. Now we can define the Feynmann propagator by

\begin{eqnarray}
\mathcal{D}(x_1-x_2, y_1-y_2, z_1-z_2, t_1-t_2)=-i\langle 0''|\mathcal{Z}\phi(x_1,y_1,z_1,t_1)\phi(x_2,y_2,z_2,t_2)|0'\rangle
\label{eqn:67}
\end{eqnarray}

where $\mathcal{Z}$ denotes z-ordering, i.e, the field with the higher z value appears on the left. In the next step, I will assume that $z_1>z_2$ and set $x_2=y_2=z_2=t_2=0$ using translational symmetry. I will also drop the subscript 1 on the former coordinates. Now the propagator becomes

\begin{eqnarray}
-i\langle 0''|\phi(x,y,z,t)\phi(0)|0'\rangle=-i\int \frac{d^3k'_1d^3k'_2}{(2\pi)^3~2\sqrt{|\lambda_1\lambda_2|}} \langle 0''| \times ~~~~~~~~~~~~~~~~ \nonumber \\ \nonumber \\
\bigg[ a'(\overrightarrow{k'_1})\exp{i(\lambda_1 z+k_{x1}x+k_{y2}y-k_{t2}t)}+\overline{a'} (\overrightarrow{k'_1})\exp{-i(\lambda_1 z+k_{x1}x+k_{y2}y-k_{t2}t)} \bigg] \nonumber \\ \nonumber \\
\times \bigg[ a'(\overrightarrow{k'_2})+\overline{a'} (\overrightarrow{k'_2}) \bigg] |0'\rangle~~~~~~~~~~~~~~~~~~~~~~~~~~~~~~~~~~~~
\label{eqn:68}
\end{eqnarray}

Simplifying this using the definitions of $|0'\rangle$ and $|0''\rangle$ and the commutation relations in Equation \ref{eqn:64} results in

\begin{eqnarray}
-i\int \frac{d^3k'}{(2\pi)^3~2\lambda}\exp{i(\lambda z+k_xx+k_yy-k_tt)}
\label{eqn:69}
\end{eqnarray}

This is the expression for $z_1>z_2$, i.e. $z>0$. Doing the same calculation for the other case and combining the results,

\begin{eqnarray}
-i\int \frac{d^3k'}{(2\pi)^3~2\lambda} \bigg[ \exp{i(\lambda z+k_xx+k_yy-k_tt)} \Theta(z)+ \nonumber \\ \nonumber \\
\exp{-i(\lambda z+k_xx+k_yy-k_tt)} \Theta(-z) \bigg]
\label{eqn:70}
\end{eqnarray}

By changing integration variables, this becomes

\begin{eqnarray}
-i\int \frac{d^3k'}{(2\pi)^3} \exp{-i(k_xx+k_yy-k_tt)} \bigg[ \frac{ e^{i\lambda z} \Theta(z)+e^{-i\lambda z} \Theta(-z) }{2\lambda} \bigg]
\label{eqn:71}
\end{eqnarray}

Now the definition of $\lambda$ (Equation \ref{eqn:51}) defines it only upto an overall sign. If we adopt the following sign convention for $\lambda$

\begin{itemize}
\item When $\lambda$ is real, it is positive
\item When $\lambda$ is imaginary it has a complex phase angle of $+\pi/2$
\end{itemize}

and if we replace the $m^2$ in Equation \ref{eqn:51} by $m^2-i\epsilon$ then the term in the square brackets can be written as

\begin{eqnarray}
-i\int \frac{dk_z}{2\pi} \frac{\exp(-ik_zz)}{k_z^2-\lambda^2}
\label{eqn:72}
\end{eqnarray}

Substituting, the propagator becomes

\begin{eqnarray}
-\int \frac{d^4k}{(2\pi)^4} \frac{ \exp{-i(k_xx+k_yy+k_zz-k_tt)} }{ k_x^2+k_y^2+k_z^2-k_t^2+m^2-i\epsilon}
\label{eqn:73}
\end{eqnarray}

which is the same as the Feynmann propagator in the usual Lagrangian formalism. Fourier transforming, we get the familiar expression for the momentum space propagator

\begin{eqnarray}
\frac{1}{ k_t^2-k_x^2-k_y^2-k_z^2-m^2+i\epsilon }
\label{eqn:74}
\end{eqnarray}

Note that the $i\epsilon$ prescription implies that $\lambda$ always has a positive imaginary part. 

\section{The Hamiltonian-like operator is non-hermitian}
Equation \ref{eqn:66} implies that $\overline{a}(\overrightarrow{k})|0'\rangle$ is an eigenstate of $H'$ with eigenvalue $\lambda$. But since $\lambda$ can take imaginary values, this means $H'$ is not hermitian. Since many important relations in the usual Hamiltonian formalism are derived using the hermiticity of the Hamiltonian, there will be some differences in how the calculations are done in our new formalism. I will try to derive the differences here. \\

First, the expression for $H'$ in terms of the ladder operators (Equation \ref{eqn:65}) and its commutation relations with the ladder operators (Equation \ref{eqn:66}) and the hermitian conjugation relation in Equations \ref{eqn:59} and \ref{eqn:60} imply that $H'$ commutes with its hermitian conjugate and is hence diagonalisable. This means the z-evolution operator $\exp{-iH'z}$ is also diagonalisable. \\

Next note that the operators $\phi$ and $\Pi$ are hermitian at $z=0$. They will not remain hermitian at other values of z because 

\begin{eqnarray}
\phi(z)=\exp(iH'z)\phi(0)\exp(-iH'z)
\label{eqn:75}
\end{eqnarray}

and this is not equal to its hermitian conjugate if $H'$ is not hermitian. But it is easy to see that it still has real eigenvalues. To see this, start from the eigenvalue equation for $\phi(0)$

\begin{eqnarray}
\hat{\phi}(0)|\phi\rangle=\phi|\phi\rangle ~~~~~~~~~~~~~~~~~~~~~~~~~ \nonumber \\ \nonumber \\
\therefore \exp(i\hat{H'}z)\hat{\phi}(0)\exp(-i\hat{H'}z)\exp(i\hat{H'}z)|\phi\rangle=\phi\exp(i\hat{H'}z)|\phi\rangle
\label{eqn:76}
\end{eqnarray}

(I am inserting the hats on operators again to distinguish them from their eigenvalues. I am also suppressing the arguments of $\phi$ other than z for brevity.) This shows that if $|\phi\rangle$ is an eigenvector of $\hat{\phi}(0)$ with a real eigenvalue $\phi$ then $\exp(iH'z)|\phi\rangle$ is an eigenvector of $\hat{\phi}(z)$ with the same real eigenvalue. Operators such as these that are not equal to their hermitian conjugate but still have real eigenvalues are called pseudo-hermitian operators. \\

Pseudo-hermitian operators in the context of quantum physics have been studied in several works such as \cite{mostafazadeh2010pseudo} and \cite{jones2005pseudo} but these works deal with pseudo-hermitian Hamiltonians, whereas we are now dealing with a non-hermitian Hamiltonian (having non-real eigenvalues) that causes the operators for other observables to evolve into pseudo-hermitian operators. \\

It should also be noted that in the case of pseudo-hermitian matrices, the left-eigenvector is not the same as the right-eigenvector but both eigenvalues must be the same. To see this, note that Equations \ref{eqn:75} and \ref{eqn:76} imply that 

\begin{eqnarray}
\langle\phi| \exp(-i\hat{H'}z)\exp(i\hat{H'}z)\hat{\phi}(0)\exp(-i\hat{H'}z)=\phi\langle\phi |\exp(-i\hat{H'}z)
\label{eqn:77}
\end{eqnarray}

So the left-eigenvector of $\hat{\phi}(z)$ with eigenvalue $\phi$ is $\langle\phi| \exp(-i\hat{H'}z)$ which is not the hermitian conjugate of the right-eigenvector. \\

Given that the z-evolution of operators in the Heisenberg picture is given by Equation \ref{eqn:75} which is the same as in the hermitian Hamiltonian case, let us now ask ourselves if the z-evolution of states in the Schrodinger picture are also the same as in the hermitian case. That is, is the state at an arbitrary value of z given by $|n(z)\rangle=\exp(-i\hat{H'}z)|n(0)\rangle$? If we assume that it is, then the expectation value of an observable of a state $|n\rangle$ will be given by

\begin{eqnarray}
\langle n(z)|\hat{\phi}|n(z)\rangle=\langle n(0)|[\exp(-i\hat{H'}z)]^\dagger \hat{\phi}\exp(-i\hat{H'}z)|n(0)\rangle  \nonumber \\ \nonumber \\
=\langle n(0)|\exp(i\hat{H'}^\dagger z)\hat{\phi}\exp(-i\hat{H'}z)|n(0)\rangle
\label{eqn:78}
\end{eqnarray}

which is not the same as what Equation \ref{eqn:75} predicts. So the assumption cannot be correct. The only way to get around this problem is by adopting the following convention: When the Hamiltonian is non-hermitian, the state of a system is described by not one but two vectors $|n_R(z)\rangle$ and $|n_L(z)\rangle$. These two vectors are taken to be equal at a particular value of z, which we can take to be $z=0$. That is,

\begin{eqnarray}
|n_R(0)\rangle=|n_L(0)\rangle
\label{eqn:79}
\end{eqnarray}

and they evolve in z in two different ways:

\begin{eqnarray}
|n_R(z)\rangle=\exp(-i\hat{H'}z)|n_R(0)\rangle  \nonumber \\ \nonumber \\
|n_L(z)\rangle=\exp(-i\hat{H'}^\dagger z)|n_L(0)\rangle
\label{eqn:80}
\end{eqnarray}

(Note the hermitian conjugation sign in the expression for $|\phi_L(z)\rangle$.) The expectation value of an operator $\phi$ in the Schrodinger picture is then defined as

\begin{eqnarray}
\langle n_L(z)|\hat{\phi}|n_R(z)\rangle=\langle n_L(0)|\exp(i\hat{H'}z)\hat{\phi}\exp(-i\hat{H'}z)|n_R(0)\rangle
\label{eqn:81}
\end{eqnarray}

That is, it is defines as a matrix element with $n_R$ on the right and $n_L$ on the left, hence the use of R and L as subscripts. And now we have the same results in the Schrodinger and Heisenberg pictures.

\medskip
\bibliography{sample}
\end{document}